\documentclass[a4paper,11pt]{article}
\ifdefined\pdfoutput\pdfoutput=1\fi
\usepackage{bm,xcolor}
\usepackage[normalem]{ulem}
\usepackage{jheppub}
\allowdisplaybreaks[1]
\makeatletter\gdef\@fpheader{}\makeatother
\graphicspath{{figures/}{../wronskian_inverse/figures/}}

\newcommand{\Wh}{\hat{W}}
\newcommand{\Fsq}{F^{2}}
\newcommand{\Tc}{T_{c}}
\newcommand{\tzero}{\tau_{c}(0)}
\newcommand{\Mgen}{M_{\rm gen}}

\begin{document}

\title{Fast holographic inversion of superconducting domes}

\author[a]{Sejin Kim}
\affiliation[a]{Center for AI and Natural Sciences, Korea Institute for
Advanced Study, Seoul 02455, Korea}
\emailAdd{sejin@kias.re.kr}

\abstract{
A holographic superconductor whose scalar mass depends on the gauge field strength, $M(\Fsq)$, reproduces a superconducting dome for a suitable $M$, but recovering $M$ from a dome has taken days.
We obtain the condition that fixes the critical temperature from two integrations, started at the horizon and at the boundary, and its derivative with respect to $M$ from the same two solutions, so training needs no finite differences and an inversion takes ten minutes to an hour.
With this speed we study the interval between the horizon value of $\Fsq$ at the lowest doping and $\Fsq=0$, which no horizon reaches and at whose end the scalar mass $M(0)$ fixes the dimension of the dual operator.
Holding $M(0)$ at several values, the pinned masses, and retraining the rest, we find reconstructions that share the minima of $M$ and agree wherever the horizons reach, apart from the deepest end, but differ on that interval.
A rule that keeps the reconstruction with the simplest closed form recovers the scalar mass and the mass function of a dome generated from a quadratic $M$.
On Gaussian and double-Gaussian domes and the measured phase diagrams of YBa$_{2}$Cu$_{3}$O$_{y}$ and 2M-WS$_{2}$ its choice rests on ties or narrow margins, and it recovers the scalar mass of only two of the four domes generated from the closed forms it keeps for them.
The dome thus constrains $M$ where its horizons reach but not the scalar mass, which a second observable would have to fix.
}

\maketitle

\section{Introduction}

Holographic duality~\cite{Maldacena97,GKP98,Witten98} turns a strongly coupled phase transition into a boundary-value problem for classical fields in one more dimension, and the holographic superconductor~\cite{Gubser08,HHH08a,HHH08b} is its standard example.
Below a critical temperature $\Tc$ a charged black hole in anti-de Sitter (AdS) space acquires charged scalar hair, and the region in which such hairy black holes exist is the superconducting phase~\cite{Hartnoll09lec,Herzog09,Horowitz10lec,CaiRev15}.

$\Tc$ is the temperature at which the linearized scalar first admits a normalizable mode regular at the horizon, the condensation mode.
Measured phase diagrams of many superconductors are domes, in which the critical temperature rises and then falls with doping~\cite{Keimer15}.
Holographic models produce domes from a modulated chemical potential~\cite{Ganguli:2013}, from competing orders~\cite{KiritsisLi:2015}, also with broken translations~\cite{Baggioli:2015dome}, from a doping field that shifts the effective mass of the order parameter~\cite{Chen:2016} and from a coupling between two charges~\cite{CaiSin:2020}, and a dome was obtained by letting the scalar mass depend on the local field strength, $M(\Fsq)$~\cite{Seo:2023,Kim:2024}.

The difficulty is finding such a holographic model.
A bulk theory that reproduces an observed phenomenon has been found by trial and error, a coupling chosen by hand, the phase diagram computed and compared with the measured one, and the coupling adjusted, so the model is tuned to the known result.
Solving the inverse problem with neural networks removes the tuning of the coupling by hand.
Given a dome, the critical temperature against doping, find an $M(\Fsq)$ that produces it, so that the bulk interaction is fixed by reproducing the measurement rather than chosen by the modeler.

Recovering bulk data from boundary observables has its own history.
Metrics have been recovered from geodesic or two-point data~\cite{Hammersley06,Bilson08}, from extremal surfaces~\cite{HubenyRT12,EngelhardtHorowitz16} and from entanglement entropy~\cite{Spillane16}.
More recently the bulk has been represented by a neural network, a composition of affine maps and pointwise nonlinearities adjusted until the output matches data~\cite{Jeong:2025,Park:2022}, or produced directly from boundary data by a generative model trained on many geometries~\cite{Park:2023,Kim:2025}.
The radial evolution of a bulk field was interpreted as the layers of a deep network~\cite{Hashimoto18,Hashimoto18b}, recast as a Boltzmann machine~\cite{Hashimoto19} and as a neural ordinary differential equation~\cite{HashimotoODE20}, and used to learn spatial geometry from entanglement~\cite{You18}, the metric and dilaton of a holographic model of quantum chromodynamics from meson spectra~\cite{Akutagawa:2020}, black-hole metrics from the shear viscosity~\cite{Yan:2020} and a bulk spacetime from the frequency-dependent conductivity~\cite{Hashimoto21}.

Recently the geometry and gauge field of a charged black hole have been reconstructed from fermionic spectral functions~\cite{Hashimoto:2026,Xiao:2026}, the latter applied to the strange-metal phenomenology of a cuprate, the metric from entanglement entropy and Wilson loops~\cite{Filev:2026} and the bulk scalar potential from the equation of state~\cite{Bea:2024,Jimenez:2026}, and each of these reconstructions meets a degeneracy or a region of the bulk that the boundary data do not probe.

The dome itself has been inverted once~\cite{Kim:2024}, with a physics-informed neural network (PINN)~\cite{Raissi19,Karniadakis21}, one whose objective is the residual of the governing equation rather than a fit to labeled examples.
The mass function it finds is lowest at the field strengths that correspond to the top of the dome, since a more negative effective mass near the horizon raises the critical temperature.
Its computational cost limits what it can be used for.
The source $\mathcal{J}$ is extracted by fitting the near-boundary expansion of a solution integrated from the horizon, and the objective built from it is minimized over roughly $10^{3}$ iterations of Adam~\cite{Kingma14}, a first-order method with per-coordinate adaptive steps.
Each iteration takes a finite-difference gradient, $n_{p}+1$ bulk integrations per point of the dome for $n_{p}$ parameters, so a single reconstruction is already expensive, and any number training does not determine has to be chosen once and then kept.

We change the forward computation first.
The solution $\psi_{H}$ regular at the horizon is integrated from the horizon and the normalizable solution $\psi_{B}$ from the boundary, both are stopped at an interior point, and their Wronskian $W$ is computed there.
On shell $W$ does not depend on where it is evaluated, and it vanishes exactly when the regular solution is normalizable, so $W=0$ is the condition $\mathcal{J}=0$, obtained without fitting the near-boundary expansion.
Each point of the dome fixes a background, and $W$ evaluated on it is that point's residual, the number training drives to zero.

The backward computation then reuses the two forward solutions.
In the radial coordinate $z$, running from the boundary at $z=0$ to the horizon at $z=1$, a Green's identity gives, for any parameter $\theta_{j}$ of $M$ that leaves $M(0)$ fixed,
\begin{equation}
\frac{\partial W}{\partial\theta_{j}}
=\int_{0}^{1}\psi_{H}(z)\,\psi_{B}(z)\,
\frac{\partial M}{\partial\theta_{j}}\big(\Fsq(z)\big)\,\frac{dz}{z^{4}} ,
\label{eq:kernel}
\end{equation}
an integral of the product of the two forward solutions already computed, with $\psi_{H}$ normalized by its value at the horizon and $\psi_{B}$ by its leading coefficient at the boundary.
The whole Jacobian, the matrix of derivatives of the residuals with respect to the parameters, is therefore obtained together with the residual at any number of parameters.
With an analytic Jacobian, exact where the residuals vanish (Appendix~\ref{app:num}), the natural method is not gradient descent but Levenberg--Marquardt~\cite{Levenberg44,Marquardt63}, which for a sum of squares builds a local quadratic model from the residual sensitivities and solves for the step.
At the hundred points per dome used here an inversion then takes from about ten minutes to an hour instead of days.

A faster inversion does not remove the part of $M$ that the dome leaves undetermined.
Every dome starts at a finite doping, and between the value $\Fsq$ takes at the horizon for that doping and $\Fsq=0$ lies an interval that contains no horizon value of the dome and on which the residual depends only weakly.
On that interval the dome does not determine $M$.
What is undetermined there is a function and not a number, and its value at $\Fsq=0$ is the scalar mass $M(0)$, the value of $M$ at the AdS boundary, which fixes the dimension $\Delta$ of the dual operator through $\Delta(\Delta-3)=M(0)$.
Section~\ref{sec:blind} calls that interval the blind region.
A single reconstruction has some function there, determined by the penalty terms of the loss, the basis and the starting point rather than by the dome.

The faster inversion makes $M(0)$ something one can vary rather than something one must choose.
At this speed $M(0)$ can be held at several values $m_{0}$, which we call pinned masses, everything else retrained at each, and the whole family inspected.
One pinned mass is natural, $m_{0}=-3/4$.
From $m_{0}=-3/4$ upward the scalar respects the stability bound of the near-horizon AdS$_{2}$ region and no critical temperature survives at zero doping, and we place the pinned masses there, at $m_{0}=-3/4$, $-1/2$ and $-1/4$.

Choosing among the members of the family needs a rule of its own.
Each reconstruction is fitted by closed forms from a fixed library, and among the reconstructions within $5\%$ of a parabola across the blind region the one reproduced with the fewest symbolic parameters, the numbers a closed form needs, is kept.
This is a restricted form of symbolic regression, the search for a closed-form expression that reproduces a curve~\cite{SchmidtLipson09,Udrescu20}, with a description-length preference for the shortest~\cite{Rissanen78}.

On every dome the reconstructions at the three pinned masses share the positions and depths of the minima of $M$ and agree over the range its horizons reach, apart from the deepest end, while they differ in the blind region (Sec.~\ref{sec:results}), so the dome constrains $M$ there but not its value at $\Fsq=0$.
On a dome solved forward from a quadratic $M$ with $M(0)=-1/2$, which plays no part in the conventions of the rule, the rule picks the reconstruction held at $-1/2$ and returns the quadratic itself (Sec.~\ref{sec:known}).
On the Gaussian and double-Gaussian domes and on the measured phase diagrams of YBa$_{2}$Cu$_{3}$O$_{y}$ (YBCO) and 2M-WS$_{2}$ it keeps one pinned mass per target (Sec.~\ref{sec:results}), but each is decided by a tie, by one symbolic parameter or by the $5\%$ threshold, so the scalar mass of these four targets is left open.
On domes generated from the closed forms the rule keeps for these targets, it returns the generating scalar mass only twice, each time by breaking a tie (Sec.~\ref{sec:results}).
Fixing the scalar mass needs a second observable, which Sec.~\ref{sec:discussion} discusses.

\section{Setup}
\label{sec:setup}

We work with a charged scalar on the Reissner--Nordstr\"om (RN) black hole in AdS$_{4}$ with two linear axions~\cite{AndradeWithers13}, the background of earlier holographic superconductors with momentum relaxation~\cite{AndradeGentle:2014,KimKimPark:2015,Baggioli:2015phases} and of the first dome inversion~\cite{Kim:2024}.
The scalar carries charge $q$, and its mass is a function of the invariant $\Fsq\equiv F_{\mu\nu}F^{\mu\nu}$, so that $M(\Fsq)|\psi|^{2}$ is a nonlinear interaction between the superconducting order and the charge density~\cite{Seo:2023,Kim:2024}.
With Newton's constant set by $16\pi G_{N}=1$ and unit AdS radius the action is
\begin{equation}
S=\int d^{4}x\sqrt{-g}\,\Big[R+6-\tfrac14\Fsq
-\tfrac12\sum_{a=1}^{2}(\partial\chi_{a})^{2}
-|D\psi|^{2}-M(\Fsq)|\psi|^{2}\Big],
\label{eq:action}
\end{equation}
with $\chi_{a}$ the axion fields, $D_{\mu}=\nabla_{\mu}-iqA_{\mu}$ and $q=1$ throughout, so the $M$ an inversion returns is the one that goes with this charge.

The field equations of Eq.~\eqref{eq:action} are
\begin{equation}
\begin{gathered}
\begin{aligned}
R_{\mu\nu}-\tfrac12Rg_{\mu\nu}-3g_{\mu\nu}
&=\tfrac12\big(1+4M'(\Fsq)|\psi|^{2}\big)F_{\mu\rho}F_{\nu}{}^{\rho}
\\
&\quad{}+\tfrac12\sum_{a=1}^{2}\partial_{\mu}\chi_{a}\partial_{\nu}\chi_{a}
+\tfrac12\big[D_{\mu}\psi(D_{\nu}\psi)^{*}+D_{\nu}\psi(D_{\mu}\psi)^{*}\big]
\\
&\quad{}-\tfrac12g_{\mu\nu}\Big[\tfrac14\Fsq+\tfrac12\sum_{a=1}^{2}(\partial\chi_{a})^{2}
+|D\psi|^{2}+M(\Fsq)|\psi|^{2}\Big],
\\
\nabla_{\mu}\Big[\big(1+4M'(\Fsq)|\psi|^{2}\big)F^{\mu\nu}\Big]
&=iq\big[\psi^{*}D^{\nu}\psi-\psi(D^{\nu}\psi)^{*}\big],
\end{aligned}
\\
\nabla_{\mu}\nabla^{\mu}\chi_{a}=0,
\qquad
D_{\mu}D^{\mu}\psi-M(\Fsq)\,\psi=0,
\end{gathered}
\label{eq:eom}
\end{equation}
with $M'=dM/d\Fsq$.

The dome is determined in the regime $|\psi|\ll1$.
At $\Tc$ the condensate is infinitesimal and its contributions to Eq.~\eqref{eq:eom} drop out at linear order, so the background is the solution with $\psi=0$.
In the radial coordinate $z$, with the AdS boundary at $z=0$ and the horizon at $z=1$, it is the RN black hole with axions~\cite{AndradeWithers13},
\begin{equation}
\begin{gathered}
ds^{2}=\frac{1}{z^{2}}\Big[-h\,dt^{2}+\frac{dz^{2}}{h}
+dy_{1}^{2}+dy_{2}^{2}\Big],
\qquad
A=Q(1-z)\,dt,
\qquad
\chi_{a}=\kappa y_{a},
\\
h=(1-z)\Big[1+z+\Big(1-\frac{\kappa^{2}}{2}\Big)z^{2}-\frac{Q^{2}}{4}z^{3}\Big],
\end{gathered}
\label{eq:rnbg}
\end{equation}
at temperature $T=(3-\kappa^{2}/2-Q^{2}/4)/4\pi$, with $Q$ the charge density.
The axions, linear in the boundary coordinates $y_{a}$, break translations homogeneously and relax momentum at a rate set by $\kappa$~\cite{AndradeWithers13}.
Here they serve two purposes only, to supply the scale $\kappa$ in which doping and temperature are measured below and, at zero charge, an extremal horizon at $\kappa^{2}=6$ whose near-horizon AdS$_{2}$ region bounds the pinned masses from below in Sec.~\ref{sec:blind}, and no transport property is used.

Near the boundary the asymptotic expansion of each field defines a source and a response.
The scalar behaves as
\begin{equation}
\psi=\mathcal{J}z^{3-\Delta}+\mathcal{O}z^{\Delta}+\dots,
\qquad
\Delta(\Delta-3)=M(0),
\label{eq:asymptotic}
\end{equation}
so that $\mathcal{J}$ is the source of the dual operator of dimension $\Delta$ and $\mathcal{O}$ is proportional to its expectation value.
Only the mass at $\Fsq=0$ enters $\Delta$, because $\Fsq$ vanishes at the boundary.
A superconducting instability is a normalizable mode, $\mathcal{J}=0$ with $\mathcal{O}\neq0$, and $\Tc$ is the highest temperature at which one exists.

On this background the scalar obeys
\begin{equation}
\Big(\frac{h}{z^{2}}\,\psi'\Big)'
+\Big[\frac{q^{2}A_{t}^{2}}{z^{2}h}-\frac{M(\Fsq)}{z^{4}}\Big]\psi=0.
\label{eq:zeromode}
\end{equation}
For two solutions $\psi_{1}$ and $\psi_{2}$ the Wronskian and its derivative, in which the terms in $\psi_{1}'\psi_{2}'$ cancel, are
\begin{align}
&W=\left(\frac{h}{z^{2}}\right)\left(\psi_{1}\psi_{2}'-\psi_{1}'\psi_{2}\right),
&W'=\psi_{1}\left(\frac{h\psi_{2}'}{z^{2}}\right)'-\psi_{2}\left(\frac{h\psi_{1}'}{z^{2}}\right)'.
\end{align}
Equation~\eqref{eq:zeromode} makes both terms of $W'$ equal to $\big(M(\Fsq)/z^{4}-q^{2}A_{t}^{2}/(z^{2}h)\big)\psi_{1}\psi_{2}$, so $W'=0$ and $W$ is independent of $z$.
Regularity at the horizon fixes
\begin{equation}
\frac{\psi'}{\psi}\bigg|_{z\to1}=-\frac{M(\Fsq_{h})}{4\pi T},
\label{eq:horizon}
\end{equation}
with $\Fsq_{h}$ the horizon value of $\Fsq$.

The phase diagram is drawn in the dimensionless plane $(x,\tau)=(Q/\kappa^{2},T/\kappa)$, charge density and temperature in units of the momentum-relaxation scale, with $x$ playing the role of doping, and a dome is the phase boundary $\tau_{c}(x)=\Tc/\kappa$, the curve below which the scalar condenses.
Because $\tau$ decreases monotonically in $\kappa$ at fixed $x$, each point of a dome fixes $(\kappa,Q)$ uniquely and independently of $M$, before any training.

The one property of Eq.~\eqref{eq:rnbg} that the rest of the paper relies on is the range of $\Fsq$.
On each background $\Fsq=-2Q^{2}z^{4}$ runs from $0$ at the boundary to $\Fsq_{h}=-2Q^{2}$ at the horizon, so a point of the dome at doping $x$ probes $M$ on $[\Fsq_{h},0]$ and nowhere else.
At the low temperatures of a dome $\kappa$ is close to its extremal value, at which $Q=(\sqrt{1+12x^{2}}-1)/x$ grows with $x$, so $|\Fsq_{h}|$ grows with doping and points at higher doping probe $M$ at more negative $\Fsq$.
The horizon values of a dome fill the horizon range $[\Fsq_{\min},\Fsq_{\max}]$, with $\Fsq_{\max}=-2Q_{\min}^{2}$ from the smallest doping and $\Fsq_{\min}=-2Q_{\max}^{2}$ from the largest.
No $\Fsq_{h}$ falls in $(\Fsq_{\max},0]$, although on every background $\Fsq$ takes all values in that interval between the horizon and the boundary.
The four domes inverted in Sec.~\ref{sec:results} are, in order, a Gaussian and a double Gaussian specified directly in $(x,\tau)$ and the measured phase diagrams of YBCO and 2M-WS$_{2}$.

\section{The blind region}
\label{sec:blind}

Every dome we invert starts at a finite doping, and below its smallest doping there are no data.
Since $\Fsq_{h}=-2x^{2}\kappa^{4}$ grows in magnitude with doping, the missing dopings are exactly the ones whose $\Fsq_{h}$ would have fallen in $(\Fsq_{\max},0]$, and $\Fsq=0$ is the end point of that interval.
Finer sampling within the measured range does not change this.

The residual depends on $M$ in that interval through only a few combinations of its values, for a reason other than the missing data.
No $\Fsq_{h}$ falls in it, so $M$ there never enters the horizon condition Eq.~\eqref{eq:horizon} and affects the residual only through the bulk, where the kernel of Eq.~\eqref{eq:kernel} weights a change of $M$ at radius $z$ by $\psi_{H}\psi_{B}/z^{4}$.
When the condensation mode exists, $\psi_{H}$ is proportional to $\psi_{B}$, which falls as $z^{\Delta}$ at the boundary, so the weight falls as $z^{2\Delta-4}$ toward $z\to0$, the boundary, where $\Fsq=-2Q^{2}z^{4}$ vanishes on every background.

Per unit $\Fsq$, the variable of the ansatz, the weight does the opposite, since $dz\propto|\Fsq|^{-3/4}\,d\Fsq$ turns $z^{2\Delta-4}dz$ into $|\Fsq|^{(2\Delta-7)/4}\,d\Fsq$, which grows toward $\Fsq=0$ with the same power on every background, so near $\Fsq=0$ the points of the dome differ mainly in a prefactor and fix only a few combinations of $M$.
We call the interval $(\Fsq_{\max},0]$ the blind region.

Weak sensitivity here does not mean that $M(0)$ has no effect.
The same number enters Eq.~\eqref{eq:asymptotic} and sets the exponent $\Delta$ with which the integration from the boundary starts, so changing $M(0)$ changes the whole dome, and the rest of $M$ has to be retrained to restore it.
The rest of $M$ can restore it, because a change at $\Fsq=0$ is absorbed by coefficients in the blind region.
The dome does not determine $M(0)$.

We therefore hold the scalar mass at each pinned mass $m_{0}$ and retrain everything else, and the family of reconstructions is the object of study.
Holding $M(0)$ exactly, rather than penalizing its distance from a value, is what makes that family well defined, and Sec.~\ref{sec:network} builds the identity $M(0)=m_{0}$ into the ansatz.

Without a pinned mass, the scalar mass that training returns is set by something other than the data.
In the Gaussian basis of Ref.~\cite{Kim:2024}, which does not fix $M(0)$ and in which every basis function is nonzero at $\Fsq=0$, the returned scalar mass follows the penalty on $M(0)$, the number of basis functions and the starting value, and without a penalty it can fall to the AdS$_{4}$ Breitenlohner--Freedman bound $-9/4$~\cite{Breitenlohner82}, below which $\Delta$ ceases to be real.

Through Eq.~\eqref{eq:asymptotic} each pinned mass fixes the dimension of the dual operator, $\Delta=\tfrac32+\sqrt{\tfrac94+m_{0}}$, which gives it a meaning independent of the training.
The three pinned masses used throughout, $m_{0}=-1/4$, $-1/2$ and $-3/4$, correspond to $\Delta=2.914$, $2.823$ and $2.725$.
What bounds the pinned masses below is a statement about zero doping, which training never evaluates.

As $x\to0$ the charge $Q=x\kappa^{2}$ vanishes, and with it $A_{t}$ and $\Fsq$ at every $z$, so Eq.~\eqref{eq:zeromode} reduces to a neutral scalar with the single constant mass $m_{0}$.
The critical temperature at $x\ll1$ is therefore a function of $m_{0}$ alone, up to corrections that vanish with $x$, and we write it $\tzero$.
Since $\tzero$ cannot separate one target from another, it is a property of $m_{0}$ and not a test of the training.

\begin{figure}[t]
\centering
\includegraphics[width=0.86\textwidth]{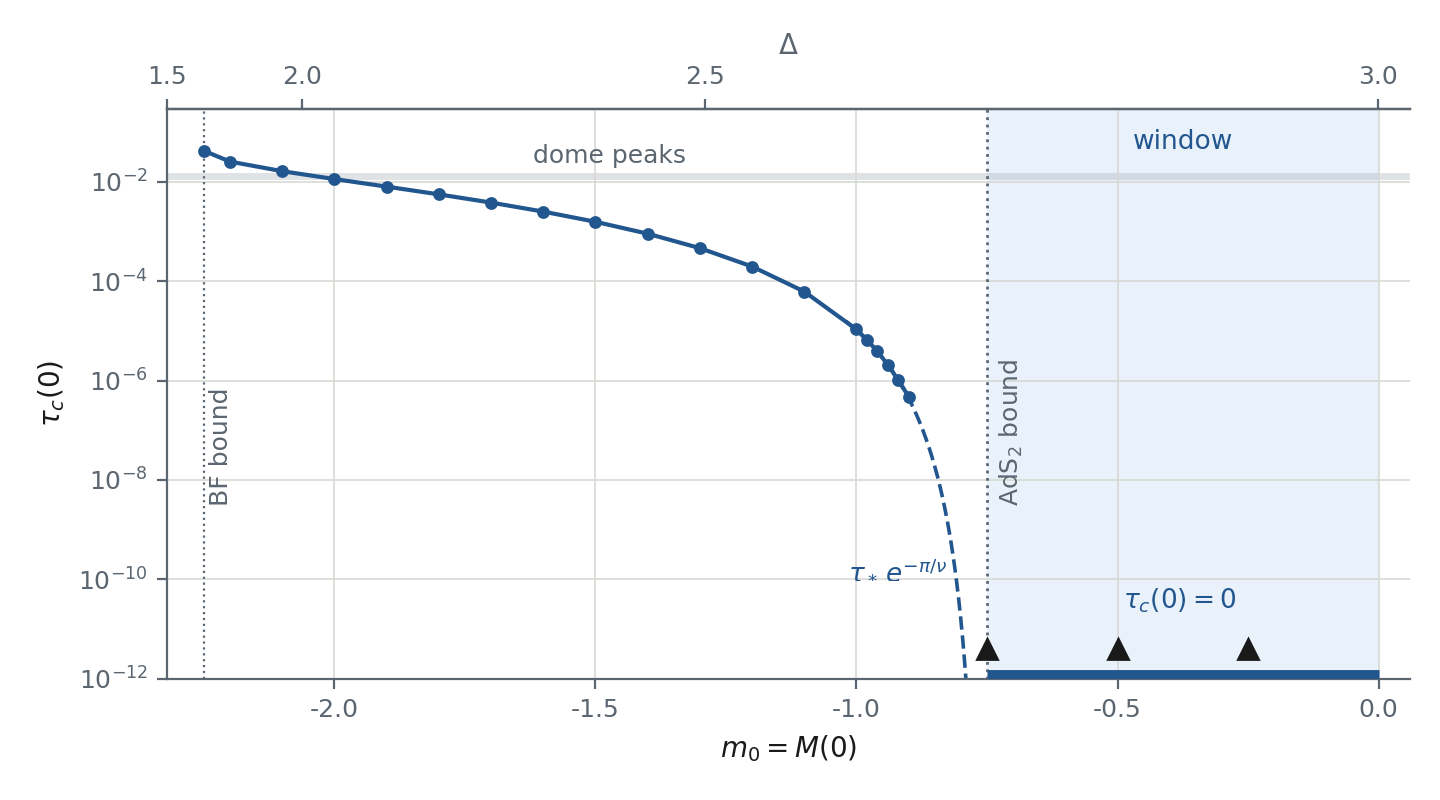}
\caption{The critical temperature at zero doping, $\tzero$, against $m_{0}$, with $\Delta$ on the top axis.
Points are $\tzero$ at $x=10^{-3}$ for constant $M=m_{0}$, and the dashed line is $\tau_{*}\exp(-\pi/\nu)$.
The shaded window holds the pinned masses (triangles), and the gray strip spans the dome peaks.}
\label{fig:tzero}
\end{figure}

The near-horizon geometry determines where $\tzero$ vanishes.
At strictly zero doping the lowest available temperature is the extremal one, $\kappa^{2}=6$, where $h\to3(1-z)^{2}$, the near-horizon geometry is AdS$_{2}\times\mathbb{R}^{2}$ with radius $L_{2}^{2}=1/3$, and substituting $\psi\sim(1-z)^{\alpha}$ in Eq.~\eqref{eq:zeromode} gives
\begin{equation}
3\alpha(\alpha+1)=m_{0} .
\label{eq:ads2}
\end{equation}
Above the AdS$_{2}$ stability bound $m_{0}L_{2}^{2}\ge-1/4$~\cite{Denef09,FaulknerLiuMcGreevyVegh}, that is for $m_{0}\ge-3/4$ or $\Delta\ge2.725$, the exponent $\alpha$ in Eq.~\eqref{eq:ads2} is real, no condensation mode exists as $x\to0$, and $\tzero=0$.
Below the bound $\alpha$ is complex and $\tzero$ is not zero.
It rises continuously and by orders of magnitude as $m_{0}$ falls, with the essential singularity expected where an AdS$_{2}$ bound is crossed~\cite{Jensen10,Iqbal11}, $\tzero\simeq\tau_{*}\exp(-\pi/\nu)$ with $\nu^{2}=-\tfrac14-\tfrac13m_{0}$ and $\tau_{*}\simeq0.55$.
Figure~\ref{fig:tzero} shows $\tzero$ against $m_{0}$.

A critical temperature at zero doping is therefore absent only for $m_{0}\ge-3/4$.
We take the window $-3/4\le m_{0}<0$, with the pinned masses $-3/4$, $-1/2$ and $-1/4$, so that no reconstruction has a critical temperature at zero doping.
The upper end keeps $\Delta$ below the marginal value $3$, which $m_{0}=0$ reaches, so that the order parameter stays a relevant operator.
The dome of Sec.~\ref{sec:known}, the one whose generating $M$ is known, has $M(0)=-1/2$, the middle pinned mass.

\section{PINN with a Kolmogorov--Arnold network}
\label{sec:model}

\subsection{The network}
\label{sec:network}

The earlier inversion~\cite{Kim:2024} is a PINN in which $M$ is expanded in Gaussian functions, the residual of each point of the dome is the source $\mathcal{J}$, and the governing equation is imposed by integrating it rather than at collocation points, the sampled points at which a PINN ordinarily penalizes the residual of the equation.
Ours keeps that structure and replaces two of its parts, the residual by the Wronskian of each point and the Gaussian expansion by a Kolmogorov--Arnold network (KAN)~\cite{Liu:2024kan}, a network whose trainable functions are expansions in B-splines, which makes the model a physics-informed KAN~\cite{Shukla:2024}.

The forward pass, the evaluation of the output from the inputs, is two integrations of Eq.~\eqref{eq:zeromode}, the branches, one from the horizon with Eq.~\eqref{eq:horizon} and one from the boundary with $\psi\sim z^{\Delta}$, which meet at an interior point.
Each is a continuous-depth network whose depth coordinate is $z$ and whose hidden state is the pair $(\psi,\psi')$, of width two because the equation is second order.
It is a residual network, whose layers each add a correction to their input, with all layers drawing their weights from the same parameters and the step chosen by the integrator.
The background $(\kappa_{i},Q_{i})$ of the $i$-th point of the dome is a fixed input, determined in Sec.~\ref{sec:setup} and never trained.

The output is produced by a head with no parameters, the final layer that maps the state to the output.
Each branch is fixed only up to an overall scale, since Eq.~\eqref{eq:zeromode} is linear and homogeneous, so a difference of the two would change under a rescaling of either.
What is invariant is whether the phase-space vectors $(\psi_{H},\psi_{H}')$ and $(\psi_{B},\psi_{B}')$ are parallel, which they are exactly when a normalizable mode regular at the horizon exists.
At the matching point $z_{\rm mid}$ the head forms
\begin{equation}
\Wh=\frac{\psi_{H}\psi_{B}'-\psi_{H}'\psi_{B}}
{\big\|(\psi_{H},\psi_{H}')\big\|\,\big\|(\psi_{B},\psi_{B}')\big\|}
\bigg|_{z_{\rm mid}}\in[-1,1],
\label{eq:what}
\end{equation}
the sine of the angle between the two terminal states.
The normalization lets the points of a dome be compared although their raw amplitudes differ by many orders of magnitude, and since $W$ is independent of $z$ on shell, the point where $\Wh$ vanishes does not depend on $z_{\rm mid}$, which we set to $0.5$.

The ansatz for $M$ is the smallest KAN, a single edge from $\Fsq$ to $M$~\cite{Liu:2024kan}, whose trainable function is one linear layer acting on fixed features,
\begin{equation}
M(\Fsq)=m_{0}+\sum_{j=1}^{100}\theta_{j}
\big[B_{j}(\Fsq)-B_{j}(0)\big],
\label{eq:basis}
\end{equation}
where the features $B_{j}$ are cubic B-splines, piecewise cubics with two continuous derivatives, on knots uniform over $[\Fsq_{\min},0]$.
The features are not learned.
The only trained quantities are the coefficients $\theta$ of that layer, and because they enter $M$ linearly the residual stays close to linear in the parameters over a step, which is why Levenberg--Marquardt applies at all.
Here, $m_{0}$ is the bias, the constant term of the layer, and the pinned mass.
The subtraction makes $M(0)=m_{0}$ an identity in the coefficients rather than a constraint on them, so no combination of the $\theta_{j}$ changes the scalar mass, and $m_{0}$ is carried as the first of the $101$ parameters and held fixed.

The features have to be localized, and the reason is the ordering of Sec.~\ref{sec:setup}.
Each feature $B_{j}-B_{j}(0)$ is nonzero on at most four knot intervals, with one exception, the last, $B_{100}-1$, which vanishes at $\Fsq=0$ and equals $-1$ below the final knot interval, so it shifts $M$ on the whole horizon range relative to the pinned mass and is the column through which a change at $\Fsq=0$ affects the horizon range.
Since a point of the dome probes $M$ on $[\Fsq_{h},0]$ and $|\Fsq_{h}|$ grows with doping, a feature placed at some $\Fsq$ affects only the points whose $\Fsq_{h}$ is more negative than that value.
The points at higher doping can therefore be matched by features the points at lower doping do not depend on, so matching them leaves the fit at lower doping unchanged.

An ordinary neural network in place of Eq.~\eqref{eq:basis} breaks this property.
Its features, the hidden units, are affine functions of $\Fsq$ passed through a pointwise nonlinearity such as $\tanh$, and each varies over the whole range of $\Fsq$.
A change made to match a high doping therefore changes $M$ everywhere, including near $\Fsq=0$ where the points at lower doping constrain it.
The Green's identity itself holds for any parameterization, so with global features this property is lost while the derivative is unaffected.

The ansatz enters the branches in three places, and two of them are easy to miss.
It sets the coefficient $M(\Fsq(z))/z^{4}$ that multiplies the state at depth $z$ in Eq.~\eqref{eq:zeromode}, which is the layer weight.
It fixes the horizon branch's input through $M(\Fsq_{h})/4\pi T$ in Eq.~\eqref{eq:horizon}.
And the boundary branch starts from $z^{\Delta}$ with $\Delta=\Delta(m_{0})$, so its input depends on the bias alone.
The pinned mass of Sec.~\ref{sec:blind} and the input of one branch are the same quantity, a second reason to hold it exactly.

\subsection{Training}
\label{sec:training}

In deep learning the backward pass, the computation of the gradient of the output with respect to every parameter, has always been the expensive part, in memory more than in arithmetic, and a research subject of its own.
Automatic differentiation evaluates the exact derivative of a computation by applying the chain rule to each of its elementary steps.
Backpropagation~\cite{Rumelhart86}, the reverse mode of automatic differentiation~\cite{Baydin18}, gives the full gradient at a small multiple of the cost of the forward pass but has to retain or recompute every intermediate state, which has led to schemes that store a few states and recompute the rest~\cite{Chen16}.
For a continuous-depth network the adjoint method~\cite{Chen18} avoids the storage by integrating an adjoint equation backward in depth, but requires a second integration whose accuracy has to be controlled.

Here the backward pass needs no integration of its own.
In network terms, Eq.~\eqref{eq:kernel} says that the gradient with respect to the weight at depth $z$, the forward activation there times the adjoint, the sensitivity of the output to the state at $z$, is $\psi_{H}(z)\psi_{B}(z)$.
Equation~\eqref{eq:zeromode} is its own adjoint, so the adjoint of the boundary branch solves the same equation with the horizon branch's values at the matching point as terminal data, which makes it the horizon branch itself continued past $z_{\rm mid}$, and conversely.
The adjoint of one branch is the forward solution of the other, and there is no separate backward integration because nothing is left to compute.

Two qualifications apply.
The first is that only $M$ is trained, entering Eq.~\eqref{eq:zeromode} as the mass of the scalar, while the branches are exact integrations of that equation for the current $M$ and have no weights of their own.
The Green's identity holds because the branches solve that equation exactly and would fail for trained approximations of them, so we do not call the branches neural networks and use network terms for them only to compare their structure with the earlier PINN.
The second concerns cost.
The residual needs each branch only as far as $z_{\rm mid}$, but the integral in Eq.~\eqref{eq:kernel} runs over the whole bulk, so a Jacobian evaluation doubles the integration.
The gain is that the observable is obtained without a fit and that the Jacobian is available at any number of parameters, not that the forward pass is shorter.

Because the ansatz is linear in $\theta$, the Jacobian factorizes.
With the product $\psi_{H}\psi_{B}$ computed once for the $i$-th point, the derivative
\begin{equation}
\frac{\partial W_{i}}{\partial\theta_{j}}
=\int_{0}^{1}\psi_{H}\psi_{B}\,\big[B_{j}\big(\Fsq(z)\big)-B_{j}(0)\big]\,\frac{dz}{z^{4}}
\end{equation}
is one contraction of a design matrix, the features evaluated on the integration grid and independent of $\theta$, against a per-point kernel.
Table~\ref{tab:cost} compares the cost of one update in the two methods.
The earlier one integrates once per point for each evaluation of its objective and, having no backward pass, needs $n_{p}+1$ evaluations to finite-difference a gradient, so its cost grows with the parameter count, while ours needs two integrations per point at any count, since the residual and the whole Jacobian are computed from the same pair of solutions.

\begin{table}[t]
\centering
\begin{tabular}{lrrr}
\hline\hline
parameters & method & integrations & seconds\\
\hline
$25$  & earlier & $728$   & $12.6$\\
      & this    & $56$    & $1.27$\\
$50$  & earlier & $1{,}428$ & $25.6$\\
      & this    & $56$    & $1.33$\\
$101$ & earlier & $2{,}856$ & $53.9$\\
      & this    & $56$    & $1.35$\\
\hline\hline
\end{tabular}
\caption{One update on the phase boundary of Sec.~\ref{sec:known} at $28$ points, with the same basis and integration accuracy.
The earlier method fits $\mathcal{J}$ from one solution per point and takes a finite-difference gradient~\cite{Kim:2024}, and ours integrates two solutions per point and uses Eq.~\eqref{eq:kernel}.}
\label{tab:cost}
\end{table}

The two methods also need different numbers of updates.
At $101$ parameters on the phase boundary of Table~\ref{tab:cost}, Levenberg--Marquardt converges in nine updates and $35$ seconds, while Adam takes about $10^{3}$ iterations~\cite{Kim:2024}, which at the last row of the table is most of a day.
Every result below uses $100$ points per dome, and there one reconstruction takes from about ten minutes to an hour, while $10^{3}$ Adam iterations at the cost per integration of Table~\ref{tab:cost} would take more than two days.
That is what makes three pinned masses on each of five phase boundaries practical.

The loss, minimized over the parameters $\theta$, is
\begin{equation}
\mathcal{L}(\theta)=\sum_{i}\Wh_{i}^{2}
+\lambda^{2}\big\|D^{(2)}\theta\big\|^{2}
+\rho^{2}\mathcal{N}^{2}\big\|C\theta\big\|^{2},
\label{eq:loss}
\end{equation}
with $i$ running over the points of the dome.
The first term is the misfit to the data.
The second is a Tikhonov term, a quadratic penalty that favors the smoothest $M$ the data allow, in which $D^{(2)}\theta$ is the second difference of $M$ on a uniform grid in $\Fsq$,
\begin{equation}
\big(D^{(2)}\theta\big)_{a}=\frac{M(u_{a+1})-2M(u_{a})+M(u_{a-1})}{(u_{a+1}-u_{a})^{2}},
\label{eq:d2}
\end{equation}
with $u_{a}$, $a=0,\dots,249$, evenly spaced over $[\Fsq_{\min},0]$.
Since $M$ is linear in $\theta$, $D^{(2)}$ is a fixed matrix, and the term penalizes curvature everywhere.
It fixes the directions the kernel leaves unconstrained.
The third term acts on the blind region alone.

That third term is the shape requirement.
Since the dome constrains $M$ only weakly in the blind region, we require of a reconstruction there not that it be flat but that it continue from the horizon range without additional structure, staying close to a parabola with a single curvature.
It is imposed by a penalty rather than a constraint because  a nearly linear $M$ fits the data and an exactly linear one does not.

The matrix $C$ takes the second divided difference of the coefficients in the blind region, each placed at the average of its three interior knots,
\begin{equation}
(C\theta)_{j}=\frac{1}{\xi_{j+1}-\xi_{j-1}}\left(\frac{\theta_{j+1}-\theta_{j}}{\xi_{j+1}-\xi_{j}}-\frac{\theta_{j}-\theta_{j-1}}{\xi_{j}-\xi_{j-1}}\right),\qquad
\xi_{j}=\tfrac13\big(t_{j+1}+t_{j+2}+t_{j+3}\big),
\label{eq:C}
\end{equation}
for every three adjacent coefficients $\theta_{j-1},\theta_{j},\theta_{j+1}$ whose B-splines reach the blind region, with $t_{j}$ and $t_{j+4}$ the end knots of $B_{j}$, so that $C\theta=0$ exactly when $M$ is linear there and the penalty is minimized by a straight line, the zero-curvature limit of a parabola.

Where the knots are uniform, $(C\theta)_{j}$ is exactly $M''(\xi_{j})/2$, half the curvature at $\xi_{j}$.
The prefactor $\mathcal{N}=\ell_{\rm b}^{2}/(4\,\delta M_{\rm b}\sqrt{n_{\rm b}})$, with $\ell_{\rm b}$ the width of the blind region, $\delta M_{\rm b}=m_{0}-M(\Fsq_{\max})$ its total rise and $n_{\rm b}$ the number of rows of $C$, makes $\rho$ the weight on the relative deviation from a straight line, since a curvature $M''$ displaces the midpoint of a segment of length $\ell_{\rm b}$ from the straight line by $M''\ell_{\rm b}^{2}/8$.
Without it the same $\rho$ would act with very different strengths on the four targets.
On a given dome $\delta M_{\rm b}$ grows with $m_{0}$, so the same $\rho$ allows more absolute curvature at the less negative pinned masses.

Both weights are set by shape and neither depends on the dome.
Every reconstruction below minimizes Eq.~\eqref{eq:loss} with the same optimizer and the same weights, $\lambda=7\times10^{-4}$ and $\rho=1$, so the shape requirement is tuned to no reconstruction.
The normalization needs $M(\Fsq_{\max})<m_{0}$, since $\mathcal{N}\propto1/\delta M_{\rm b}$.

\subsection{The rule}
\label{sec:rule}

Choosing a value inside the window needs a rule that the residual does not provide.
The rule applies two criteria to the family, and the first tests the shape requirement, which the loss imposes only as a penalty.
For each reconstruction we compute its deviation from a parabola over the blind region,
\begin{equation}
    \epsilon_{2}=\frac{\max\big|M-\sum_{r=0}^{2}\beta_{r}(\Fsq)^{r}\big|}{|m_{0}-M(\Fsq_{\max})|},\qquad
    \beta=\arg\min\int_{\Fsq_{\max}}^{0}\Big(M-\sum_{r=0}^{2}\beta_{r}(\Fsq)^{r}\Big)^{2}d\Fsq,
\end{equation}
with $\beta=(\beta_{0},\beta_{1},\beta_{2})$ the coefficients of the least-squares parabola in $\Fsq$, and we keep those below a threshold of $5\%$.

Under the second criterion, closed forms are fitted to every reconstruction, to the curve the spline returned rather than to the phase boundary, and among those that pass the first criterion the pinned mass kept is the one whose reconstruction is reproduced by the closed form with the fewest symbolic parameters.
The closed forms come from a fixed library,
\begin{equation}
\bar{M}=m_{0}+\sum_{r=1}^{n_{c}}c_{r}(\Fsq)^{r}
+\sum_{l=1}^{n_{f}}d_{l}\,\Phi(s_{l},w_{l}),
\qquad
\Phi(s,w)=e^{-((\Fsq-s)/w)^{2}}-e^{-(s/w)^{2}},
\label{eq:library}
\end{equation}
a polynomial of degree $n_{c}$ plus $n_{f}$ Gaussian functions of amplitude $d_{l}$, center $s_{l}$ and width $w_{l}$, with the subtraction in $\Phi$ keeping $\bar{M}(0)=m_{0}$ identically.
The degree runs over $n_{c}=1,2,3$ when $n_{f}\le1$ and over $n_{c}=2,3$ when $2\le n_{f}\le5$, and the number of symbolic parameters is $k=n_{c}+3n_{f}$, the pinned mass not being counted.

Each point of the dome depends on $M$ only between $\Fsq=0$ and its own horizon value $\Fsq_{h}$.
The end of $M$ on which only the last two points, the two highest dopings, depend is where the reconstruction is least reliable, so the closed form is fitted without it.
Every point of the dome stays in the reconstruction. The coefficients and widths are bounded and neighboring Gaussian functions are kept apart.\footnote{%
Amplitudes are bounded by $|d_{l}|\le20$, widths by $0.02\le w_{l}\le10$ and polynomial coefficients by $|c_{r}|\le100$.
The center of the first Gaussian function lies inside the comparison grid, and each further center lies at larger $\Fsq$ than the previous one, by $5\%$ to $100\%$ of the grid length, and may lie past $\Fsq=0$, where no background reaches.
}
A closed form reproduces a reconstruction when its root-mean-square curve residual over the compared interval of $\Fsq$ is at most $0.6\%$ of the range of $M$ on the whole curve, the accuracy of the reconstruction itself, and ties in $k$ between pinned masses are broken by the mean error of the phase boundary each closed form regenerates, the phase-boundary error, which is the relative difference between the regenerated and the target $\tau_{c}$ averaged over the points of the dome.

\section{Results}
\label{sec:allresults}

\subsection{A phase boundary whose mass function is known}
\label{sec:known}

Every target of Sec.~\ref{sec:results} specifies a dome and leaves $M$ unknown, so there the question can only be what the reconstruction returns and never whether it returns the $M$ that generated the dome.
Here we write one down,
\begin{equation}
\Mgen(\Fsq)=-\tfrac12+2.125\,\Fsq+0.525\,(\Fsq)^{2},
\label{eq:closure}
\end{equation}
solve Eq.~\eqref{eq:zeromode} forward at $100$ values of $x$ to obtain a phase boundary, and train on it by the same procedure, without using $\Mgen$.
The generating scalar mass is $-1/2$.
Equation~\eqref{eq:closure} is chosen to give a dome of about the right shape, not a mass function anything is known to have, so recovering it is a statement about the inversion and nothing else.

At the three pinned masses the phase-boundary errors are $0.259$, $0.247$ and $0.254\%$ from $m_{0}=-1/4$ down to $-3/4$, nearly equal although the scalar mass changes by $1/2$.
The smallest is at the generating scalar mass, but the differences reflect the weak sensitivity of Sec.~\ref{sec:blind} and do not single out a minimum.

On this dome  the part left out of the comparison is $\Fsq<-4.719$, the $\Fsq_{h}$ of the third-highest doping.
All three reconstructions pass the first criterion of Sec.~\ref{sec:rule}, and the second selects the generating scalar mass and the generating quadratic.
At $m_{0}=-1/2$ that form is the pure quadratic,
\begin{equation}
\bar{M}_{\rm known}=-\tfrac12+2.1191\,\Fsq+0.5228\,(\Fsq)^{2},
\label{eq:closed-known}
\end{equation}
against the $2.125$ and $0.525$ of Eq.~\eqref{eq:closure}.
At the other two pinned masses the form with the fewest symbolic parameters that reproduces the reconstruction includes a Gaussian function, a line with one at $m_{0}=-3/4$ and a quadratic with one at $-1/4$, so the count is two at the generating scalar mass against four and five elsewhere, with no tie.

The rule prefers the reconstruction whose shape needs the fewest parameters, and on this dome that is the one held at the generating scalar mass, because a reconstruction held elsewhere has to bend across the blind region to reach its pinned mass, and representing that bend requires additional Gaussian functions.

\begin{figure}[t]
\includegraphics[width=\textwidth]{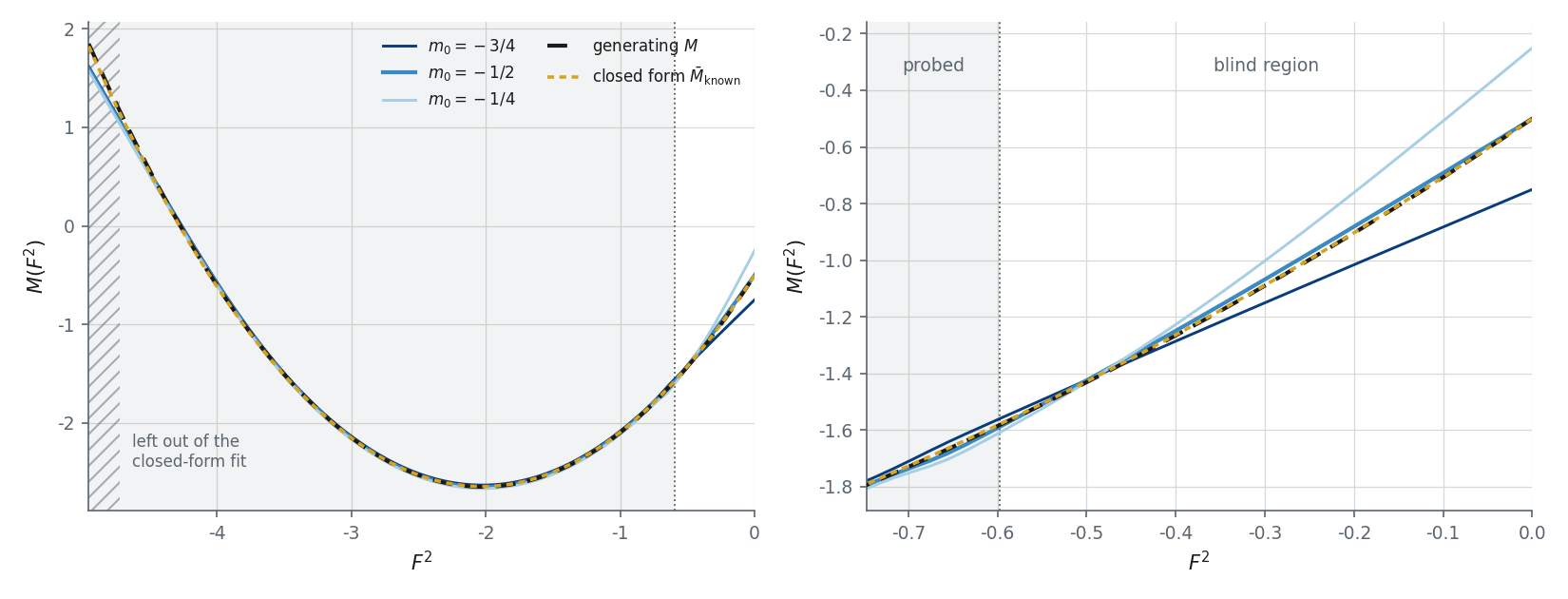}
\caption{Reconstructions from the dome of Eq.~\eqref{eq:closure} at each pinned mass, with $\Mgen$ (dashed) and $\bar{M}_{\rm known}$ (orange dotted), in full (left) and near the blind region (right).
Shading marks the horizon range, the dotted line $\Fsq_{\max}$, and diagonal lines the end left out of the closed-form fit.}
\label{fig:class}
\end{figure}
\begin{figure}[t]
\includegraphics[width=\textwidth]{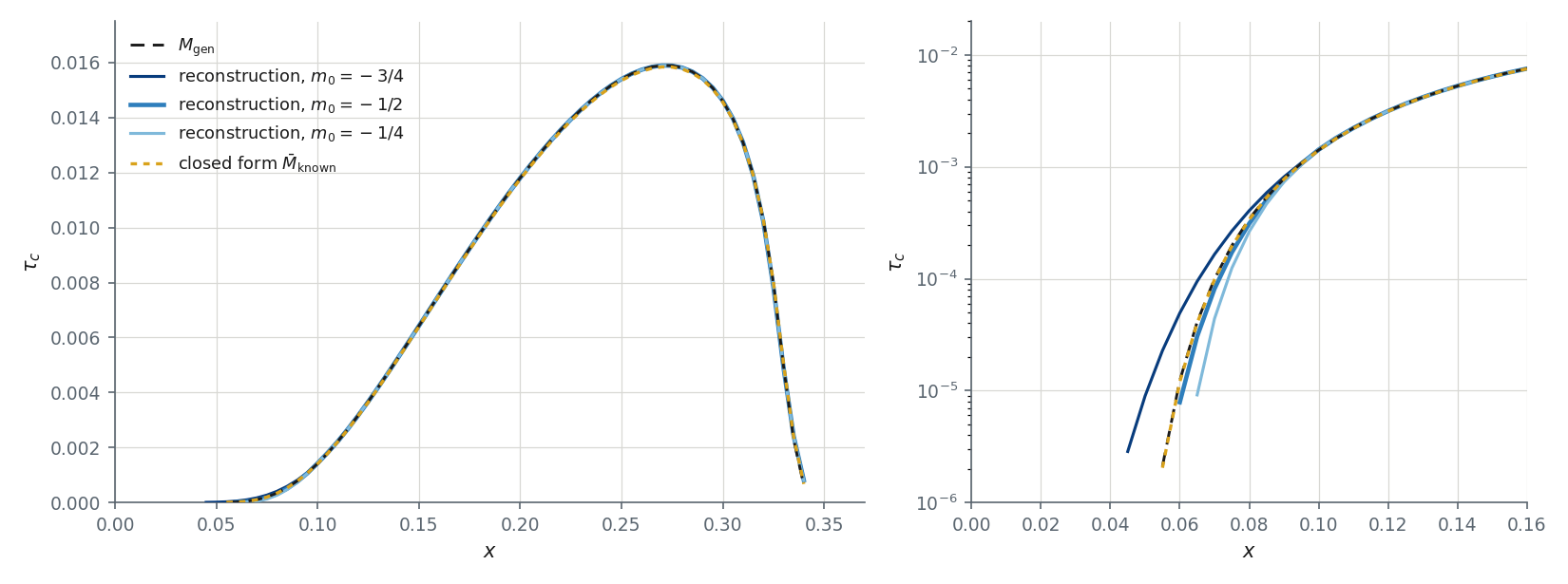}
\caption{Phase boundaries regenerated by the reconstructions of Fig.~\ref{fig:class} and by $\bar{M}_{\rm known}$, against that of $\Mgen$ (dashed), on which the $100$ points of the dome lie, on a linear scale (left) and at low doping on a logarithmic scale (right).}
\label{fig:knowndome}
\end{figure}

Figure~\ref{fig:class} shows which parts of $M$ the data constrain.
Over the horizon range, apart from the end marked with diagonal lines, the three reconstructions agree with $\Mgen$ and with each other.
Over the blind region the curves are set by the pinned mass and the shape requirement.
Only the curve held at $-1/2$ stays close to $\Mgen$, and the other two deviate from it near $\Fsq_{\max}$ and end at their own pinned masses.
The phase boundaries they regenerate, in Fig.~\ref{fig:knowndome}, lie on the phase boundary of $\Mgen$ over the range of the data and part only below the smallest doping, according to their pinned masses, while the closed form $\bar{M}_{\rm known}$ stays on the phase boundary of $\Mgen$ there too.
As in Sec.~\ref{sec:blind}, a change of $m_{0}$ is absorbed by coefficients in the blind region, which the data constrain only weakly, while for the quadratic, whose values on the horizon range depend on its value at $\Fsq=0$, changing the pinned mass changes the curve where the data are, and the fit to the data becomes worse.

In other words, identifiability, whether the data single out one $M$, is a property of the class of functions and not of the dome.
One phase boundary and one set of pinned masses leave the answer undetermined at $101$ parameters and determine it at two, and Fig.~\ref{fig:class} shows the region where the difference arises.
On the two Gaussian targets of Sec.~\ref{sec:results} the phase-boundary error changes little across the pinned masses, while on the two measured ones it grows toward the most negative pinned mass.

\subsection{Four reconstructions}
\label{sec:results}

Four domes are inverted, and training uses the phase boundary and nothing else.
Two are specified directly as domes in $(x,\tau)$, a Gaussian in $x$, which is the example of Ref.~\cite{Kim:2024}, and a sum of two Gaussians, harder because the minimum between the peaks requires compensating structure at intermediate $\Fsq$.
No $M$ is known for either.
Two are measured.
The first is the critical temperature of YBCO against hole doping $p$~\cite{Sato:2017}, the data used by the earlier method~\cite{Kim:2024}, digitized from the published figure into $34$ points.
The second is the carrier-density phase diagram of 2M-WS$_{2}$~\cite{Zhao:2021}, which peaks at $8.6$~K and whose abscissa is already a density.

The axes of the bulk theory are dimensionless, so each measured phase diagram enters with two scale factors, $x=a_{x}p$ and $\tau=a_{\tau}\Tc$ with $p$ the measured abscissa, chosen once so that the data lie in the part of the $(x,\tau)$ plane that the background covers, the material-dependent ambiguity noted in Ref.~\cite{Kim:2024}.
They are $a_{x}=1.78$ and $a_{\tau}=1.339\times10^{-4}$~K$^{-1}$ for YBCO, with $p$ the hole doping, and $a_{x}=0.2$ and $a_{\tau}=1.5\times10^{-3}$~K$^{-1}$ for 2M-WS$_{2}$, with $p$ the carrier density in units of $10^{21}$~cm$^{-3}$.
Both maps are linear through the origin, so any phase boundary below returns to measured units through $p=x/a_{x}$ and $\Tc=\tau/a_{\tau}$.
Every target enters training as $100$ points, sampled from a smoothing spline on the measured ones.

\begin{figure}[t]
\includegraphics[width=\textwidth]{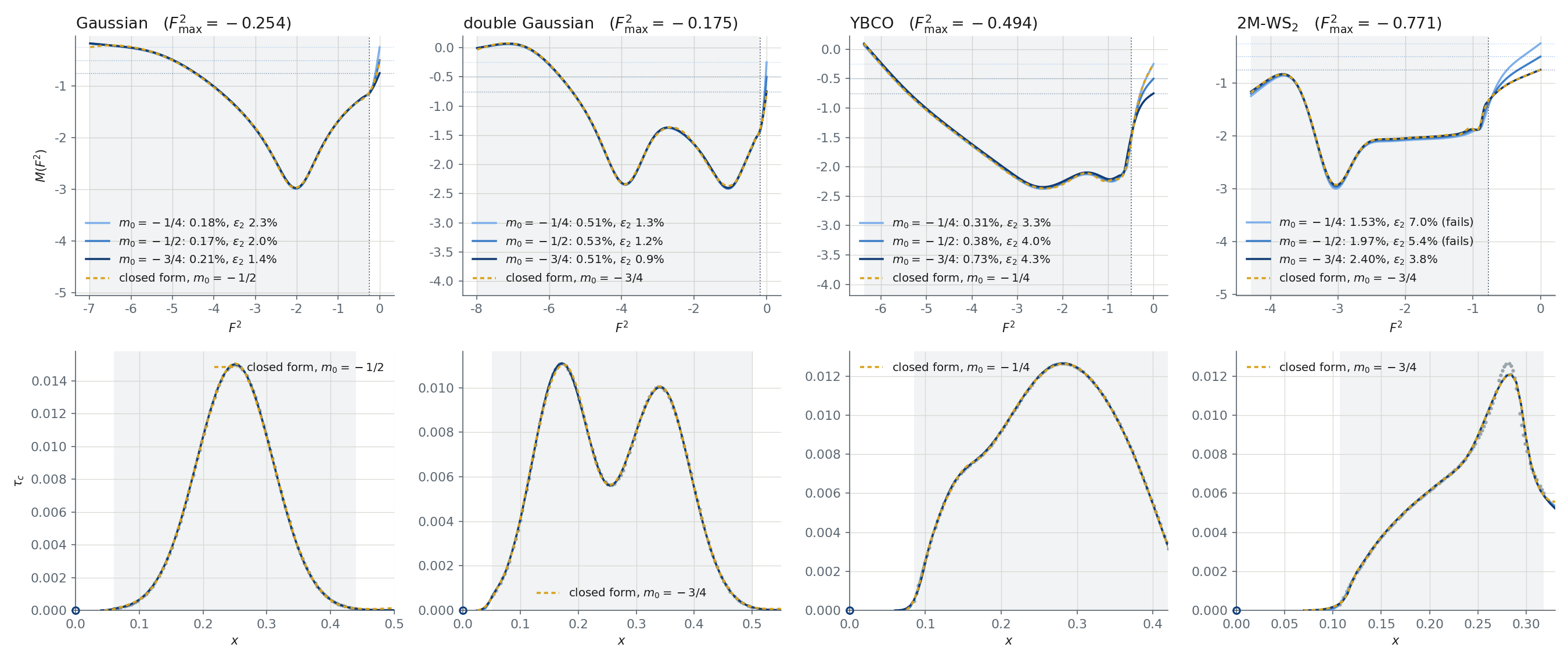}
\caption{Top, the reconstructed $M$ at each pinned mass, the closed form the rule keeps (orange), the horizon range (shaded) and $\Fsq_{\max}$ (dotted), with the phase-boundary error and $\epsilon_{2}$ in the legends.
Bottom, the regenerated $\tau_{c}(x)$ over the target (gray points), with the training range of $x$ shaded.}
\label{fig:pins}
\end{figure}

Figure~\ref{fig:pins} shows the four targets at the three pinned masses, all with the same pair of weights.
Every trained reconstruction reproduces the phase boundary of its target.
The peak is reproduced closely on three of the four, and on 2M-WS$_{2}$ it comes out a few percent low, because the reconstruction smooths the sharp maximum at $x=0.28$.

The top row shows the three reconstructions of each target in full.
The structure visible in such a plot, the positions and depths of the local minima of $M$ and the plateau on 2M-WS$_{2}$, is common to all three and is therefore a property of the dome.
The value of a curve at $\Fsq=0$ is not, being its pinned mass, and neither is its shape across the blind region, which the shape requirement determines and which is therefore an inference rather than a reconstruction.

The two criteria of Sec.~\ref{sec:rule} leave one pinned mass per target.
Every one of the twelve reconstructions rises monotonically across the blind region.
At the $5\%$ threshold on the deviation from a parabola, the two least negative pinned masses of 2M-WS$_{2}$ fail the first criterion and every other reconstruction passes.

On the four targets the second criterion again leaves out the end on which only the two highest dopings depend.
The counts are not small.
The Gaussian target needs eleven symbolic parameters at every pinned mass, the double Gaussian fourteen or fifteen, YBCO eleven at every pinned mass and 2M-WS$_{2}$ fifteen to eighteen, with eighteen, the largest form in the library, on the one reconstruction that passes the first criterion.
No pure polynomial passes on any of the twelve reconstructions.
In three of the four forms kept a center offset is at its smallest allowed value, one in the double Gaussian and YBCO forms and two in the 2M-WS$_{2}$ form, so the bound of the library, not the curve, fixes the closest pair of Gaussian functions there, and in the Gaussian form one amplitude is at its bound $|d_{l}|=20$.

The Gaussian target is thus a three-way tie, and the regenerated errors that break it differ by less than $0.01\%$, so the one kept, $-1/2$, is in effect undetermined.
YBCO is a three-way tie as well, broken at $-1/4$ by $0.662\%$ against $0.750$ and $1.325\%$, and the double Gaussian and 2M-WS$_{2}$ have no tie.

The kept closed forms regenerate the phase boundary less accurately than the reconstructions (Table~\ref{tab:select}), because a curve matched to $0.6\%$ of the range of $M$ does not reproduce the phase boundary as closely as the reconstruction does.
Table~\ref{tab:select} collects one value per target, the one the rule infers, with the dimension it implies.
Appendix~\ref{app:closed} writes the closed forms kept for the four targets out.

\begin{table}[t]
\centering
\begin{tabular}{lcccrrr}
\hline\hline
phase boundary & $m_{0}$ & $\Delta$ & $(n_{c},n_{f})$ & $k$ & spline & closed\\
\hline
generated by Eq.~\eqref{eq:closure} & $-1/2$ & $2.823$ & $(2,0)$ & $2$ & $0.247\%$ & $0.538\%$\\
\hline
Gaussian        & $-1/2$ & $2.823$ & $(2,3)$ & $11$ & $0.172\%$ & $2.286\%$\\
double Gaussian & $-3/4$ & $2.725$ & $(2,4)$ & $14$ & $0.512\%$ & $1.003\%$\\
YBCO            & $-1/4$ & $2.914$ & $(2,3)$ & $11$ & $0.306\%$ & $0.662\%$\\
2M-WS$_{2}$     & $-3/4$ & $2.725$ & $(3,5)$ & $18$ & $2.398\%$ & $2.424\%$\\
\hline\hline
\end{tabular}
\caption{The pinned mass kept for each phase boundary, its dimension, the kept form $(n_{c},n_{f})$ of Eq.~\eqref{eq:library} with $k=n_{c}+3n_{f}$, and the phase-boundary error of the $101$-parameter reconstruction and of the closed form, written out in Eqs.~\eqref{eq:closed-known} and \eqref{eq:closed-gauss} to \eqref{eq:closed-ws2}.}
\label{tab:select}
\end{table}

The rule does not always return the generating scalar mass.
On domes generated from variants of Eq.~\eqref{eq:closure}, with $M(0)$ between pinned masses or a Gaussian function added inside the blind region, it returns the generating scalar mass, or the pinned mass nearest to it, together with the quadratic.
On domes generated from the closed forms of Appendix~\ref{app:closed} at their own $m_{0}$, every pinned mass needs the same number of symbolic parameters for YBCO, the double Gaussian and the Gaussian, and the smallest count falls at a wrong pinned mass for 2M-WS$_{2}$, so the rule returns the generating scalar mass in two of the four cases, both through the tie-break.
On each of the four targets a factor of two in a threshold or another convention of the rule moves the kept pinned mass.

\section{Discussion}
\label{sec:discussion}

The value this paper reports for each target is the output of three choices, the window of Sec.~\ref{sec:blind} and the two criteria of the rule, each of which could be made differently.
There is a window, the values of $m_{0}$ at which no critical temperature survives at zero doping, set by the stability bound of the near-horizon AdS$_{2}$ region.
There is the first criterion, a bound on the deviation from a parabola across the blind region.
And there is a preference for the reconstruction reproduced by the closed form with the fewest symbolic parameters.
None of the three comes from the residual, which does not determine $M(0)$ and constrains the rest of the blind region in only a few combinations.
What the training of Sec.~\ref{sec:training} contributes is that the rule can be applied at all, to fifteen reconstructions, each fitted by the closed forms of the library.

What would falsify the rule, as opposed to disagreeing with its output, is a case where the generating $M$ is known and the rule does not return it.
Section~\ref{sec:known} is such a case, and there the rule returns it.
Such cases also include the four domes at the end of Sec.~\ref{sec:results}, whose generating $M$ are the closed forms of Appendix~\ref{app:closed}, and two of them falsify the rule in this sense.
On the four targets of Sec.~\ref{sec:results}, however, the pinned masses are decided by ties, by a single parameter or by the threshold of the first criterion, so the values kept are not evidence for their scalar masses.

A second observable could take the place of the rule.
The dome is one function of doping and leaves $M(0)$ open, but observables of the condensed phase, or of pair fluctuations above $\Tc$, depend on the dimension $\Delta$ of the dual operator, and the pair susceptibility has been proposed as a measurement of that dimension in quantum critical metals~\cite{She:2011}.
Computing one of them from the reconstructions at the different pinned masses would show how far apart they are in measurable terms.

There are three things this paper does not establish.
It does not measure any material's scalar mass, since both measured phase diagrams enter with two scale factors chosen by hand and the number reported moves when they change.
It does not claim that the mass functions displayed are unique, since the reconstructions at different $m_{0}$ nearly agree on the horizon range and differ by the full spacing of the pinned masses only in the blind region.
And it does not claim to generalize beyond the RN black hole with axions.
The construction relies on $|\Fsq_{h}|=2Q^{2}$ growing along every dome, which is why localized features are the right ansatz in Sec.~\ref{sec:network}, and a background on which $|\Fsq_{h}|$ does not grow monotonically along a dome would require that ansatz to be reconsidered.

Two assumptions about the condensed phase are not tested here.
Every $\Tc$ above is the onset of the condensation mode, which is the transition temperature only if the transition is continuous, and couplings of the order parameter to the gauge field can make it first order~\cite{Franco:2009}, with the transition temperature then above the onset.
Below $\Tc$ the coupling feeds back into the Maxwell equation through the factor $1+4M'(\Fsq)|\psi|^{2}$ of Eq.~\eqref{eq:eom}, which has to stay positive for the gauge field to keep a positive kinetic term, and that bounds the condensate where $M'<0$, at more negative $\Fsq$ than each minimum of every reconstruction.
Both questions need the hairy black holes below $\Tc$, which this paper does not construct.

Without the shape requirement, at $\rho=0$, a measured reconstruction has a shallow local minimum between $\Fsq_{\max}$ and $\Fsq=0$, and removing that minimum is what raises the phase-boundary error at the common weights.
Such a minimum requires a sufficiently wide blind region, and the geometry determines its width.
Its share of the range of $\Fsq$ is $(Q_{\min}/Q_{\max})^{2}$, so a narrower doping range gives a wider blind region.
The measured doping ranges are the narrower ones, but on the same ranges the two Gaussian domes have the minimum weakly or not at all, so a narrow range alone does not produce it, and which feature of a measured dome produces it we leave open.

In the language of inverse problems the method is a least-squares fit constrained by an ordinary differential equation.
The condition for a condensation mode is a Sturm--Liouville problem, as in analytic estimates of $\Tc$~\cite{SiopsisTherrien:2010}, its derivative Eq.~\eqref{eq:kernel} is first-order perturbation theory for a self-adjoint operator, known elsewhere as the adjoint-state method~\cite{Plessix:2006}, and which parts of $M$ the dome determines is the question of inverse Sturm--Liouville theory~\cite{Borg:1946} and of resolution analysis~\cite{BackusGilbert:1968}, to which the blind region is the answer here.
Most of the speed-up comes from the analytic Jacobian, which makes an update about $n_{p}/2$ times cheaper in integrations than a finite-difference one at the same basis and accuracy (Table~\ref{tab:cost}), and Levenberg--Marquardt reduces the number of updates further, so a finite-difference Jacobian with Levenberg--Marquardt would keep the smaller factor and lose the larger.

What the dome constrains, then, is the mass function where its horizons reach, with the same minima at every pinned mass, and not the scalar mass, which the window of Sec.~\ref{sec:blind} bounds and a second observable would have to fix.

\acknowledgments
The author is supported by a KIAS Individual Grant (AP103401) via the Center for AI and Natural Sciences at the Korea Institute for Advanced Study.
The code for the method and for the computations reported here, and the drafts of this manuscript, were written with the assistance of Claude (Anthropic).
The code and the data behind every figure are available from the author on request.
The formulation of the problem, the design of the tests and the interpretation of the results are the author's own, and the author takes full responsibility for the content.

\appendix

\section{Numerical parameters}
\label{app:num}

Integration uses DOP853, the explicit Runge--Kutta method of order eight of Dormand and Prince, with relative and absolute accuracy $10^{-11}$ and $10^{-13}$ during training, and $10^{-9}$ and $10^{-12}$ when a phase boundary is regenerated from a reconstructed $M$ by root finding.
The boundary branch starts at $z_{0}=5\times10^{-4}$ and the horizon branch at a distance $10^{-6}$ from the horizon, which the leading-order near-horizon expansion used there needs to be small compared with $4\pi T/|M(\Fsq_{h})|$.

The matching point is $z_{\rm mid}=0.5$, and the integral of Eq.~\eqref{eq:kernel} is evaluated on $1200$ points of a grid denser at both ends.
The coefficients are bounded by $5$ in absolute value, the box within which the trust-region reflective solver used for every reconstruction takes its steps~\cite{Branch99}, a Levenberg--Marquardt variant that reflects steps off the bounds and solves its subproblem by singular value decomposition with column scaling taken from the Jacobian.
Its trust region is the neighborhood of the current coefficients in which a quadratic model of the objective is trusted, and each step stays inside it.

The Jacobian of Eq.~\eqref{eq:what} keeps the kernel and drops the derivative of the normalization, which is exact at the solution because that term multiplies $\Wh$ itself.
At the common weights $\Wh$ does not vanish at the minimum, so the dropped term is of the order of $\Wh$ there.
The solver therefore stops at a stationary point of an objective that differs from Eq.~\eqref{eq:loss} by terms proportional to $\Wh$, which the same two solutions could remove, since the derivative of the normalization follows from them by variation of parameters.
The misfit is $\Wh$ and not the relative error in $\tau_{c}$, which it becomes to first order after division by $\tau_{i}\partial_{\tau}\Wh_{i}$, so the points are weighted by the sensitivity of $\Wh$ to $\tau$, and the weights $\lambda$ and $\rho$ are set by shape rather than by a discrepancy principle, which would set them so that the misfit in $\tau_{c}$ matches the noise of the data.
Every free coefficient leaves $M(0)$ fixed, 
\begin{equation}
\frac{\partial M(0)}{\partial\theta_{j}}=0,
\end{equation}
so the boundary term that a variation of $\Delta$ contributes appears only in the derivative with respect to the pinned mass, which is held fixed.

\section{The closed forms kept}
\label{app:closed}

Written out with the coefficients rounded to four significant figures, the closed forms kept for the four targets are
\begin{align}
\bar{M}_{\rm Gauss}&=-\tfrac12-1.720\,\Fsq-0.1329\,(\Fsq)^{2}
\nonumber\\*
&\quad-0.6817\,\Phi(-2.058,0.5540)+3.856\,\Phi(-0.1112,1.373)
\nonumber\\*
&\quad+20.00\,\Phi(1.174,0.7243),
\label{eq:closed-gauss}\\
\bar{M}_{\rm 2Gauss}&=-\tfrac34+1.610\,\Fsq+0.07314\,(\Fsq)^{2}
\nonumber\\*
&\quad-0.7547\,\Phi(-3.929,0.6025)-4.727\,\Phi(-3.441,2.422)
\nonumber\\*
&\quad-8.193\,\Phi(-0.4487,1.739)-0.7738\,\Phi(-0.04944,0.2911),
\label{eq:closed-dgauss}\\
\bar{M}_{\rm YBCO}&=-\tfrac14+0.9616\,\Fsq+0.1558\,(\Fsq)^{2}
\nonumber\\*
&\quad-0.8962\,\Phi(-2.077,1.525)-0.7821\,\Phi(-0.9308,0.4321)
\nonumber\\*
&\quad-0.5790\,\Phi(-0.6120,0.2209),
\label{eq:closed-ybco}\\
\bar{M}_{{\rm WS}_{2}}&=-\tfrac34+0.4336\,\Fsq+0.4112\,(\Fsq)^{2}
+0.07788\,(\Fsq)^{3}
\nonumber\\*
&\quad-1.748\,\Phi(-3.092,0.3804)-1.438\,\Phi(-2.411,0.7924)
\nonumber\\*
&\quad-0.9339\,\Phi(-1.358,0.6616)-0.1731\,\Phi(-1.143,0.1802)
\nonumber\\*
&\quad-0.3447\,\Phi(-0.9286,0.09334),
\label{eq:closed-ws2}
\end{align}
with $\Phi$ the Gaussian function of Eq.~\eqref{eq:library}, while the test of Sec.~\ref{sec:known} keeps Eq.~\eqref{eq:closed-known}.
Every amplitude $d_{l}$ in them is negative except the last two of the Gaussian form.

\bibliographystyle{JHEP}
\bibliography{references}

\end{document}